\documentclass{article}
\usepackage{spconf,amsmath,graphicx}
\usepackage{bbold}
\usepackage{xcolor}

\usepackage{booktabs,multirow}
\usepackage{bigdelim}
\usepackage{float}
\usepackage{cleveref}

\title{Identifying birdsong syllables without labelled data}
\name{Author(s) Name(s)} %\thanks{Thanks to XYZ agency for funding.}}
\address{Author Affiliation(s)}
\name{Mélisande Teng$^{1,2\hspace{0.15em}*}$ \qquad Julien Boussard$^{1,3\hspace{0.15em}*}$ \qquad David Rolnick$^{1,3}$ \qquad Hugo Larochelle$^{1,2}$}
  
  \address{$^{1}$ Mila - Quebec AI Institute,
  $^{2}$ Université de Montréal,
      $^{3}$ McGill University,  
* Equal contribution}
\begin{document}
%\ninept
%
\maketitle
\begin{abstract}
Identifying sequences of syllables within birdsongs is key to tackling a wide array of challenges, including bird individual identification and better understanding of animal communication and sensory-motor learning. 
Recently, machine learning approaches have demonstrated great potential to alleviate the need for experts to label long audio recordings by hand. However, they still typically rely on the availability of labelled data for model training, restricting applicability to a few species and datasets.  
In this work, we build the first fully unsupervised algorithm to decompose birdsong recordings into sequences of syllables. We first detect syllable events, then cluster them to extract templates --syllable representations-- before performing matching pursuit to decompose the recording as a sequence of syllables. 
We evaluate our automatic annotations against human labels on a dataset of Bengalese finch songs and find that our unsupervised method achieves high performance. We also demonstrate that our approach can distinguish individual birds within a species through their unique vocal signatures, for both Bengalese finches and another species, the great tit.  
% We demonstrate the potential of our approach as a tool to facilitate ecological studies, using the automated syllable annotations obtained with our method to identify individual birds, through their unique vocabulary and vocal signature, on previously unlabelled data. 
%Finally, we show how our approach can be a tool for facilitating faster human annotation or understanding intra vs inter-individual variability. 
\end{abstract}
\begin{keywords}
bioacoustics, audio segmentation, unsupervised learning, matching pursuit.
\end{keywords}
\section{Introduction}
% \textcolor{red}{ use some references from Elly Knight's review}https://www.sciencedirect.com/science/article/abs/pii/S0169534724001186

%\textcolor{blue}{First paragraph about bioacoustics, species classification vs individual ID / call type (cite previous ICASSP papers) Unsupervised vs supervised approach and manual labelling}

%Thanks to a recent development in sound recording technology and the possibility of recording sounds in dense, covered environments where camera traps cannot be used, bioacoustics--the study of animal sounds, a subfield of remote sensing--is becoming key for ecologists. 
With recent advances in sound recording technologies, bioacoustics, the study of animal sounds, has emerged as an important tool for conservation, in particular for birds \cite{PENAR2020100847}. Indeed, birdsong recordings are widely available and are informative for monitoring bird populations, understanding their behavior, and assessing biodiversity \cite{birdnet}. In particular, breaking down songs into their unit elements, \textit{syllables}, is relevant for many applications from studying sensory-motor learning \cite{sensory_motor_learning} to identifying individual birds \cite{repertoire_aiid} or analyzing regional dialects \cite{vocal_dialects}. 
These studies typically rely on annotations of recordings at the syllable level, obtained through manual labelling, a time- and cost-intensive process which is prone to inconsistency between annotators. Biologists delimit each syllable event manually \cite{manual_segment} in the spectrogram of a song recording, and each individual bird's recordings are usually treated independently, 
% Therefore, syllable ``a'' of an individual might not be the same as syllable ``a'' in another individual, subsequently 
posing challenges for studying the similarities between individual birdsongs for example. 

Recently, machine learning models have proven promising in bioacoustics, especially for the task of species classification where supervised deep learning approaches have achieved state-of-the-art accuracy \cite{birdnet, perch}, leveraging vast amounts of labelled data from databases such as XenoCanto \cite{xeno_canto}. However, much less labelled data is available at the individual, song or syllable level, hindering the scalability of these methods to new tasks and datasets. 

To speed up the annotation process, semi-automated pipelines and annotation tools have been developed \cite{VoICE, review_sequence_analysis}.
Cohen et al.~\cite{tweetynet} showed that a supervised neural network could segment recordings of Bengalese finches and classify syllables following human-defined labels. Alexander et al.~\cite{crypticowl} proposed to segment notes with the \textit{sci-kit maad} Python package \cite{ulloa2021scikit}, before extracting acoustic features for each of the notes and clustering them using UMAP. 
% However, 
This procedure is very sensitive to the choice of hyperparameters, requiring human supervision. 
These methods facilitate automated labelling of recordings but are limited by the extensive manual effort needed to create a training set for supervised learning. 
Indeed, the number of possible syllables can scale with the number of individual birds, meaning that the training dataset often must grow with the size of the inference dataset.

In this paper, we propose a fully unsupervised approach to (i) find templates corresponding to distinct syllable shapes, (ii) segment spectrograms of birdsongs into sequences of syllables, and (iii) label these syllables. % templates representative of the syllable shape, their frequency and their time of occurrence in the recording.
The unsupervised nature of the approach enables fast data exploration and annotation of birdsong recordings. 
% and the produced syllable sequences can then be used for applications such as individual or call type identification. 
While we focus on birds in this study, our method can be applied to other taxa. Moreover, our method allows us to extract shared structure across individuals by annotating multiple recordings together. 
 % However, such models typically compress the individual differences between songs to extract species-specific information. Many ecological tasks such as population estimation or animal tracking are based on individual identification, which is thus done by extensive manual labelling (CITE). 

%In order to automate these problems, we believe that it is first needed to extract features that represent the intra-species various sources of variability in an unsupervised way, before designing unsupervise clustering methdos that take these features as pinput to perform a wide range of tasks. In this paper, we take advantage of unsupervised signal-decomposition methods to automatically segment the bird songs into a sequence of syllables each defined by a template, representative of the syllable shape, a frequency and time. We hope that this sequence will contain useful information to help identifying call types as well as individuals. 

% In order to extract meaningful syllables from unlabelled data, 
We propose a strategy 
% that couples clustering and matching pursuit \cite{matching_pursuit} methods, 
drawing inspiration from research in spike-sorting \cite{dartsort}, where the goal is to extract neural spiking events from unlabeled, high-dimensional electrophysiological recordings  \cite{spikesorting_review}. More precisely, we run an amplitude threshold-based detection of syllable events, which we then cluster to create syllable templates. Finally, we run matching pursuit on the full recordings against our templates to reconstruct the sequences of syllables uttered by the birds. 
% We demonstrate the potential of this approach as an effective and valuable tool to facilitate the annotation process of recordings with minimal human supervision. 
% cluster these syllables using HDBSCAN on their principal components, run matching pursuit to extract more syllables, before iteratively refining the clustering and the detection of syllables. ADD SOMETHING ABOUT "developing a tool for facilitating the annotation by experimentalists i.e. minimal manual tuning"

% , and compare our performance to the supervised method TweetyNet \cite{tweetynet}, %although a supervised method that is moreover informed of the total number of clusters.  
We demonstrate that our method successfully identifies bird vocal signatures in two different contexts. First, we evaluate our method on a dataset of Bengalese finch songs annotated at the syllable level \cite{bengalesefinches} and find that it achieves high precision and recall across the 4 individuals in this dataset. We also show that our method, by identifying each bird's bag of syllables, can provide insights into the identity of each individual bird within the dataset.
Secondly, we consider a dataset of great tit recordings without labels on individual syllables \cite{great_tits}. Again, we find that our method extracts information reflecting the identity of individual birds and song types. 

%\textcolor{blue}{maybe ADD A BIT MORE ABOUT ML APPROACHES FOR AIID (Ovenbird, hierarchical AIID https://ieeexplore.ieee.org/abstract/document/10890076 etc...) + write about the different sources of variability (species, subspecies / locations, call types etc...)}

\section{Method}\label{sec:method}
% \vspace{-0.8em}

% \begin{figure}[hbt!]
% \centering
% \centerline{\includegraphics[width=0.8\columnwidth]{figures/final_methods.pdf}}
% \caption{\textbf{Illustration of our segmentation pipeline.} \textbf{A)} Each song is preprocessed as a normalized spectrogram and passed as input to our model. The spectrogram is then thresholded and syllables are detected as connected components (white boxes). \textbf{B)} Principal components are input to HDBSCAN to cluster the syllables. \textbf{C)} Templates are constructed as the cluster medians. \textbf{D)} Matching pursuit is run using the set of templates as basis functions to automatically detect and cluster the syllables in all recordings. Boxes correspond to the matching pursuit detections, with colors showing the matching to the corresponding templates in C.}
% \label{fig:pipeline}
% % \end{center}
% \end{figure}

\subsection{Detection of syllable events}
% Following the preprocessing steps outlined in \cite{great_tits}, we transform the raw audio to a Mel-spectrogram with  FFT length 512 and bandpass filter [1kHz, 10kHz], and convert it to a dB scaled spectrogram (\cref{fig:pipeline}, A).
%The spectrogram is converted to a dB scaled spectrogram, with minimum value set to 0, and thresholding the output at 65dB below the peak (\cref{fig:pipeline}, A).  
% Then, we substract to the spectrogram the minimum value of the spectrogram so that all spectrograms have minimum value zero. 

Recordings are first preprocessed into spectrograms. Then, syllable events (\textbf{SE}s) are extracted as the connected components of the spectrogram, separated by any signal lower than a threshold $\eta$ \cite{note_separation}. %(\cref{fig:pipeline}, A, white boxes)% In our work, we choose $\eta=10  \text{dB}$.% Each connected component is then considered a syllable event .

\subsection{Clustering and refinement of templates}\label{sec:initial_clustering}

The goal of the following steps is to find templates for each syllable in the ``vocabulary'' of birds in a given dataset. Ideally, a single template would be found for each syllable. First, we zero-pad each detected SE to a fixed size that reflects the maximal temporal and frequency span allowed for each SE, to obtain images of a fixed size centered on each detected SE.

Then, we fit a PCA on all detected SEs, and use HDBSCAN \cite{hdbscan} on the first 3 principal components (PCs) to produce an \textbf{initial clustering} 
% of the syllable events 
%(\cref{fig:pipeline}, C)
. We refine this clustering by \textbf{splitting} the clusters. We fit a separate PCA on each cluster of SEs and run HDBSCAN on the first 2 PCs for each cluster. % When multiple sub-clusters are found by HDBSCAN, each sub-cluster forms a new one. 
By performing PCA on each cluster separately, we capture refined low-dimensional features allowing to separate different SEs that are part of a single cluster.   
% splits of the initial clusters. 
% The goal of this step is to %capture the unique manifold structure of each cluster maximize the purity of the clusters. 

However, this can lead to over-splitting clusters (i.e. multiple clusters corresponding to the same syllable). This issue is addressed with a \textbf{merge} step. First, we generate templates by taking the median of all SEs in a cluster. The templates are designed to be representative of each cluster syllable shape, and we use the median which is more robust to outliers than the mean. We then compute the following distance for each pair of templates $T_1, T_2$:

\begin{equation}
d(T_1, T_2) = \frac{||T_1 - T_2||^2}{\max(||T_1||^2, ||T_2||^2)}
\end{equation}

This pairwise normalized distance is used to perform hierarchical clustering with complete linkage \cite{hierarchical_clustering}, using a threshold $h \in [0, 1]$. Whenever the normalized norm of the difference between multiple templates is lower than $h$ times the templates' norm, the templates are merged. This ensures that the resulting templates all encode different syllable shapes. 

\subsection{Matching pursuit and iterative template refinement}

Finally, we perform inference with a procedure inspired by \textbf{matching pursuit} \cite{matching_pursuit}. 
Given the initial set of templates obtained from the above steps, this step decomposes a given, possibly unseen recording into a set of detected SEs, all assigned to their corresponding templates, by minimizing the norm of the residual between the original recording and the sum of the templates at their corresponding detected times: 

\begin{equation}\label{deconv_objective}
D(T, t, f) = ||V - \sum_{k} S_{T_k}(t_k, f_k)||_2, 
\end{equation}

\noindent where $V$ is the original recording and $S_{T_k}(t_k)$ is a detected SE at time $t_k$ and frequency $f_k$ assigned to template $T_k$.

This matching pursuit objective is optimized using a greedy procedure, proposed in \cite{yass}. Namely, we compute, for all templates, times, and frequencies, the difference between the signal $V$ and the residual norm $D(T,t, f)$. The local maxima of this time series are considered detected SEs and are assigned to the template and frequency that maximize the value at these timesteps. The procedure finds templates, times and frequencies that lead to the highest decrease in signal norm over the entire recording. 
% This results in a 1 dimensional time series for each timestep. Each peak of the time series is considered as a detected syllable and assigned to the template and frequency that maximizes the difference between signal and residual at this timestep. 
Moreover, assuming that a bird will not sing multiple syllables at the same time, we enforce a collar around each SE to prevent overlapping of detected SEs, applying max-pooling over $D(T, t, f)$, keeping the syllables that match best with the signal if there is overlap. %To do this, we take the max-pooled time series $D(t, l, f)$, by pooling the maximum in a given interval. 

The split-merge and matching pursuit steps can then be repeated to obtain a final syllable annotation of the recordings. Indeed, matching pursuit improves the detection of SEs, and these SEs can then be used in the split-merge step, to obtain refined templates and produce improved assignments of syllables to SEs at inference time. 

\subsection{Postprocessing}

To improve the obtained sequence of SEs at inference, we remove detected SEs that are assigned to templates where the syllable signal duration is less than one timestep, as these likely correspond to noise. 

%\textcolor{blue}{TODO (maybe not here): explain how parameters are chosen / how it's truly unsupervised}

%\textcolor{blue}{Need to specify in the next section the differences (1D vs 2D) between datasets + One round of deconv vs 2 in great tits}

% \textcolor{blue}{+ Add that deconv improves upon initial clustering i.e. ``In our experiments, we found that a single iteration of split-merge iteration improved significantly on the initial clustering and gave satisfactory results overall.''}

\section{Experiments}

We apply our method on two datasets of different species: Bengalese finches (\emph{Lonchura striata domestica}) and great tits (\emph{Parus major}). Hyperparameters 
% in steps A and B in Fig. \ref{fig:pipeline} (threshold values and number of PCs) 
were kept the same for both datasets. We preprocessed recordings following \cite{bengalesefinches, great_tits} and chose the same threshold value for initial detection as \cite{great_tits}. Bengalese finch song syllables span the entire frequency spectrum. Thus, we ran matching pursuit only in the time dimension for this species. Great tits songs cover only a small frequency range, therefore, we set a box size of 100 timesteps by 100 frequency bins in the log-scale for detected SEs. 
We set parameters $\eta = 10 \text{dB}$ and $h = 0.33$, and the minimum and maximum cluster size parameters of HDBSCAN to 10 and 200 respecitvely (\cref{sec:initial_clustering}). We found that they worked well in practice and did not tune them further. 
%While no dataset-specific hyperparameter tuning was conducted, we obtained satisfactory results on both datasets. % In  the Bengalese finches dataset, ... whereas the great tits ... . Therefore, \textcolor{blue}{ specify the differences (1D vs 2D) in the deconv vs 2 in great tits}

\vspace{-0.25em}

\subsection{Bengalese Finches dataset}

\begin{figure}[hbt!]
\centering
\centerline{\includegraphics[width=0.95\columnwidth]{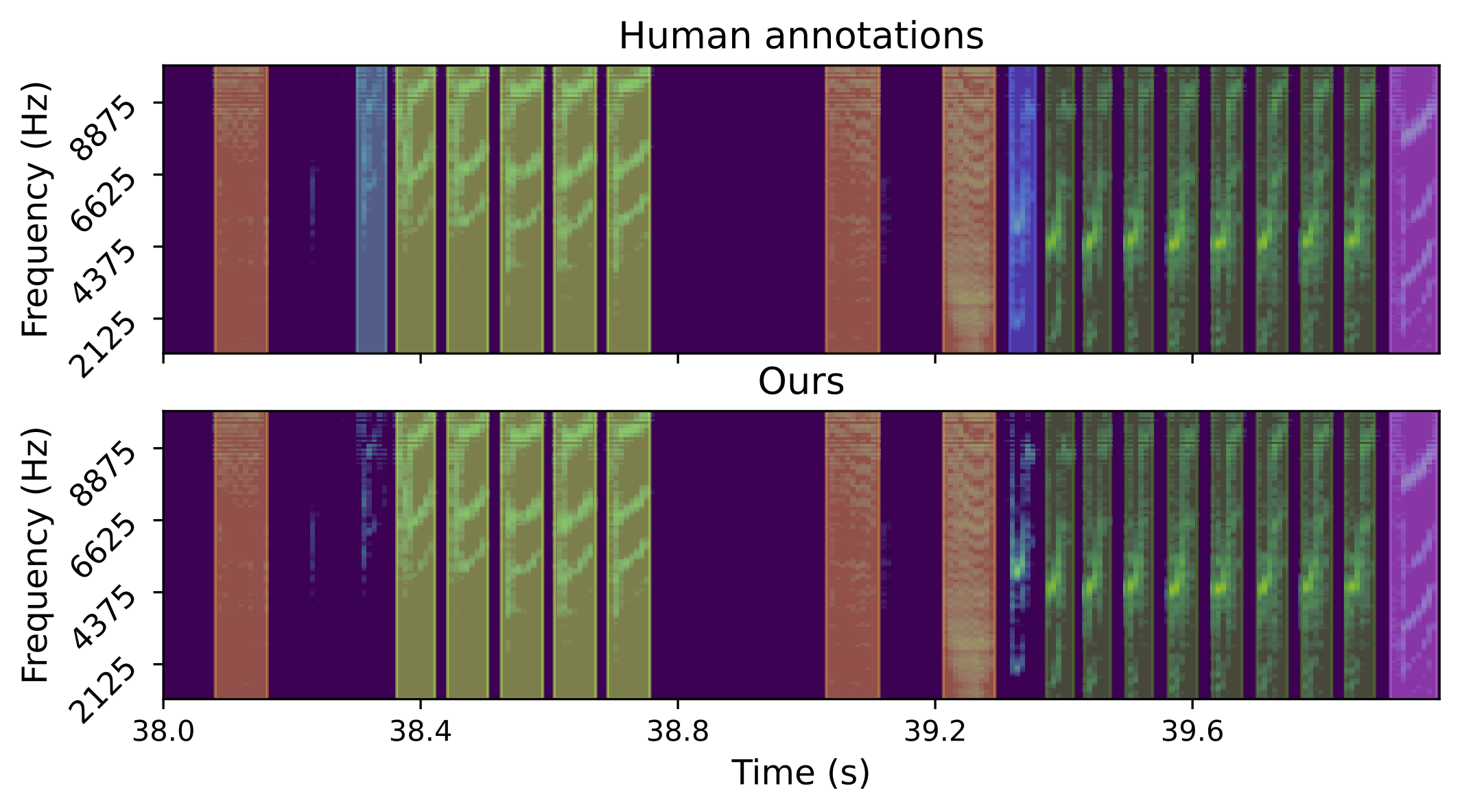}}
\vspace{-0.07in}
\caption{\textbf{Example Bengalese finch recording snippet}, with human syllable annotations (top row) and our method's output (bottom row) highlighted by the colored regions. }
\label{fig:BF_results}
% \end{center}
\end{figure}

\textbf{Dataset.} We consider the Bengalese Finches dataset \cite{tweetynet} which consists of a collection of 1.75h to 3.5h of recordings for each of the 4 individuals, manually annotated at the syllable level. Part of an example recording with corresponding human annotations is shown in the top row of \cref{fig:BF_results}.

\textbf{Setup.} We split the data into a support set, used for obtaining templates, and a query set on which matching pursuit is applied, and our method evaluated. For each individual, we sample a support set of 10 minutes of recordings, and the query set is composed of the rest of the recordings. 
% Note that all the syllables of an individual might not be represented in the support set. 
% ** should we mention that we choose 10 min/indiviudal because that is what TweetyNet does?**
We consider five different splits and report results averaged over the five query sets. We consider two setups: the \textit{single}-individual setup, where templates are obtained for each individual using only this individual's support set, and the \textit{multi}-individual setup, where templates are obtained from the combined support sets of all individuals, and shared across all individuals.  

\textbf{Evaluation.} As the labels --human syllable annotations-- are not shared across individuals, we evaluate our method on each individual separately with 1) detection precision and recall to evaluate how
many SEs are correctly detected, regardless of the class assignment, 2) 
micro-averaged  and weighted-averaged precision to account for syllable class imbalance in the bird songs, and 3) weighted--averaged recall.%,  weighted by the number of ground truth events corresponding to each syllable class. 

% While our method is fully unsupervised, we evaluate it by comparing it to the ground truth labels. 
To calculate these metrics, we proceed as follows. 
% Each template is derived from a cluster of support set syllables. 
On the support set, we assign each detected SE the label of any ground truth SE occurring at the same time, or an ``empty'' label if none. Then we take the majority label across events in each cluster of detected SEs,. This gives us a correspondence between identified and ground truth clusters. We then compute our metrics on the query set given this correspondence. 
\begin{table*}[hbt!]
\centering
\begin{footnotesize}

\resizebox{\textwidth}{!}{\begin{tabular}{|c|c|cc|cc|c|cc|}
\hline
Setup & ID & \multicolumn{2}{c|}{Detection} & \multicolumn{2}{c|}{Precision} & Recall & Median \# of& \# ground-truth \\

& & \textbf{Precision} & \textbf{Recall} & \textbf{Micro-averaged} & \textbf{Weighted-averaged} & \textbf{Weighted-averaged} & templates & syllables\\
\hline

\multirow{5}{*}{\rotatebox[origin=c]{90}{Single}} & \textbf{gr41rd51} & 0.84$\pm$0.04 & 0.66$\pm$0.03 & 0.71$\pm$0.06 & 0.43$\pm$0.11 & 0.47 $\pm$ 0.07 & 8& 26\\

& \textbf{bl26lb16} & 0.96$\pm$0.01 & 0.93$\pm$0.01 & 0.91$\pm$0.04 & 0.85$\pm$0.08 & 0.87$\pm$0.05 & 9& 20\\

& \textbf{gy6or6} & 0.74$\pm$0.01 & 0.42$\pm$0.00 & 0.97$\pm$0.01 & 0.44$\pm$0.03 & 0.44$\pm$0.01 & 15& 17\\

& \textbf{or60yw70} & 0.87$\pm$0.04 & 0.63$\pm$0.05 & 0.89$\pm$0.02 & 0.80$\pm$0.08 & 0.60$\pm$0.07 & 9& 16\\
\cline{2-9}
& \textbf{average} & 0.85 & 0.66 & 0.87 & 0.63 & 0.60 & 10& 20\\

\hline
\multirow{5}{*}{\rotatebox[origin=c]{90}{Multi}} & \textbf{gr41rd51} & 0.79 $\pm$ 0.06 & 0.66 $\pm$ 0.04 & 0.76 $\pm$ 0.04 & 0.54 $\pm$ 0.04 & 0.50 $\pm$ 0.04 & 22& 26\\
& \textbf{bl26lb16} & 0.95$\pm$0.00 & 0.79$\pm$0.03 & 0.96$\pm$0.01 & 0.86$\pm$0.07 & 0.76$\pm$0.03 & 29& 20\\
& \textbf{gy6or6} & 0.75$\pm$0.02 & 0.41$\pm$0.01 & 0.97$\pm$0.01 & 0.47$\pm$0.04 & 0.39$\pm$0.00 & 29& 17\\
& \textbf{or60yw70} & 0.77$\pm$0.06 & 0.41$\pm$0.04 & 0.93$\pm$0.02 & 0.87$\pm$0.05 & 0.40$\pm$0.04 & 26& 16\\
\cline{2-9}
& \textbf{average} & 0.82 & 0.57 & 0.91 & 0.63 & 0.69 & 26 & 20\\
\hline

\end{tabular}}
\vspace{-0.5em}
\end{footnotesize}
\caption{Bengalese finches query set results averaged on 5 random splits for each individual}
\label{tab:results}
\end{table*}

% \begin{table*}[hbt!]
% \centering
% \begin{footnotesize}

% \resizebox{\textwidth}{!}{\begin{tabular}{c|c|cc|cc|c|cc}

% Setup & ID & \multicolumn{2}{c|}{Detection} & \multicolumn{2}{c|}{Precision} & Recall & \# predicted & \# ground-truth \\

% & & \textbf{Precision} & \textbf{Recall} & \textbf{Micro-averaged} & \textbf{Weighted-averaged} & & templates & syllables\\
% \hline

% \multirow{4}{*}{\rotatebox[origin=c]{90}{Single}} & \textbf{gr41rd51} & 0.84$\pm$0.04 & 0.67$\pm$0.03 & 0.65$\pm$0.09 & 0.46$\pm$0.09 & 0.46 $\pm$ 0.06 & 9& 26\\

% & \textbf{bl26lb16} & 0.96$\pm$0.01 & 0.95$\pm$0.02 & 0.85$\pm$0.13 & 0.80$\pm$0.20 & 0.82$\pm$0.13 & 9& 20\\

% & \textbf{gy6or6} & 0.74$\pm$0.01 & 0.42$\pm$0.00 & 0.97$\pm$0.02 & 0.57$\pm$0.04 & 0.42$\pm$0.01 & 15& 17\\

% & \textbf{or60yw70} & 0.87$\pm$0.04 & 0.63$\pm$0.05 & 0.87$\pm$0.04 & 0.82$\pm$0.07 & 0.62$\pm$0.05 & 10& 16\\

% \hline
% \multirow{4}{*}{\rotatebox[origin=c]{90}{Multi}} & \textbf{gr41rd51} & 0.93 $\pm$ 0.05 & 1.00 $\pm$ 0.00 & 0.66 $\pm$ 0.03 & 0.53 $\pm$ 0.04 & 0.44 $\pm$ 0.04 & 23& 26\\
% & \textbf{bl26lb16} & 0.98$\pm$0.00 & 0.99$\pm$0.01 & 0.96$\pm$0.00 & 0.86$\pm$0.07 & 0.76$\pm$0.03 & 29& 20\\
% & \textbf{gy6or6} & 0.93$\pm$0.01 & 0.99$\pm$0.01 & 0.96$\pm$0.01 & 0.74$\pm$0.06 & 0.39$\pm$0.00 & 30& 17\\
% & \textbf{or60yw70} & 0.90$\pm$0.03 & 1.00$\pm$0.00 & 0.94$\pm$0.02 & 0.87$\pm$0.05 & 0.39$\pm$0.04 & 26& 16\\

% \end{tabular}}
% \vspace{-0.5em}
% \end{footnotesize}
% \caption{Bengalese finches query set results averaged on 5 random splits for each individual}
% \label{tab:results}
% \end{table*}

\subsection{Great Tits dataset}

\textbf{Dataset.} We consider the Great Tits dataset \cite{great_tits}, comprising 109,963 songs from 454 individuals, annotated at the song type and individual level. Song types were labelled for each individual separately and are not shared across individuals. 

\begin{figure}[hbt!]
\centering
\centerline{\includegraphics[width=0.95\columnwidth]{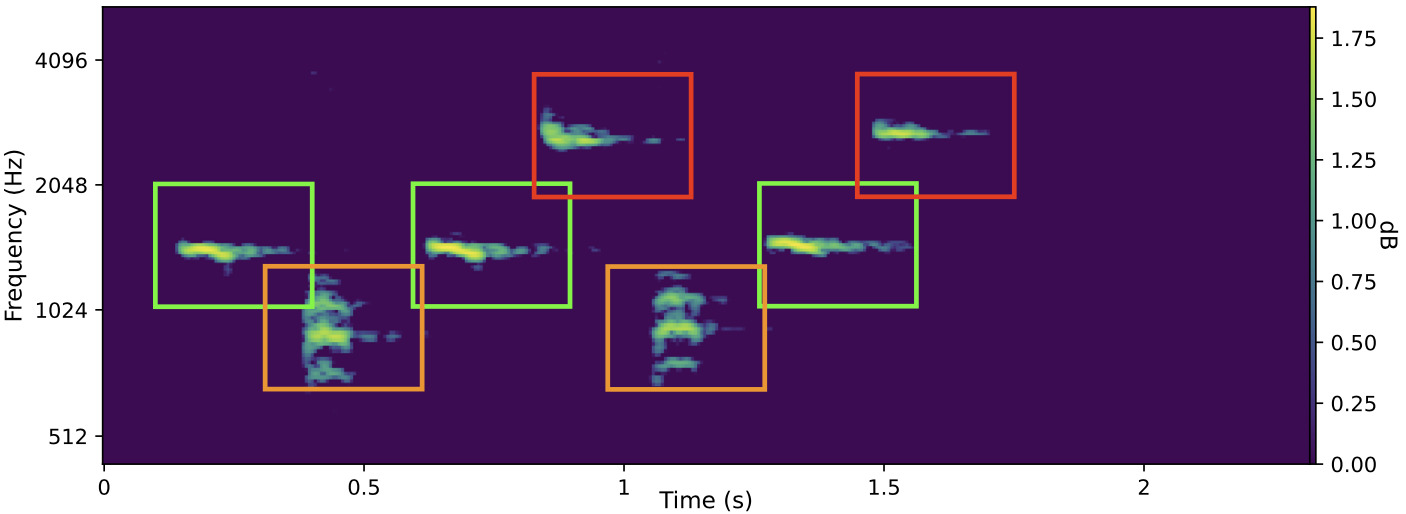}}
\caption{\textbf{Method output on one great tit song.} Detected syllable events are outlined by the colored boxes, with the colors indicating the template assignments.}
\label{fig:results_greattits}
% \end{center}
\end{figure}

\textbf{Setup.} We select randomly 2,000 songs from 25 individuals and run our algorithm on the entire collection of these 2,000 songs. An example great tit song, with syllables annotated with our method is shown in \cref{fig:results_greattits}. 

\textbf{Evaluation.} To show that our method identifies the vocal signatures of the 25 individuals, we first compute a ``bag of syllables'' (BoS) for each song. Each song consists in a set of detected syllables and their template assignment. Because templates might be matched to SEs with the same shape but very different frequency, we further bin templates by frequency, obtaining 251 ``augmented'' labels for SEs. Each event matched with a given template in a given frequency bin is counted as an occurrence in the BoS, giving us a 251-dimensional representation of each song. We compute the t-SNE 2D representation of the BoS, and verify that this 2D representation effectively separates the individuals and song types. We compare it with the 2D t-SNE embeddings of the 1028-dimensional Perch embeddings \cite{perch} of these songs.

\section{Results and discussion}
\subsection{Bengalese Finches dataset}

\vspace{-0.15in}
\vspace{-0.07in}

Table \ref{tab:results} summarizes performance metrics in the single and multi setup for each individual. Our method achieves high detection and micro-averaged precision (respectively 0.82 and 0.91 in average across individuals in the multi and 0.85 and 0.87 in the single setup). Detection and clustering recall is lower in general because our method misses low occurrence frequency syllables in the single setup, and because it tends to ``oversplit'' clusters between the many templates in the multi setup. Also for this reason, 
% We argue that, with little human supervision, it is possible to merge the split clusters and improve the recall without much loss in the precision.
precision improves in the multi setup while recall is higher in the single setup. 
% By finding templates across different individuals, different templates are more likely to reflect different syllables rather than intra-variability of syllables in an individual. 

\begin{figure}[hbt!]
\centering
\centerline{\includegraphics[width=\columnwidth]{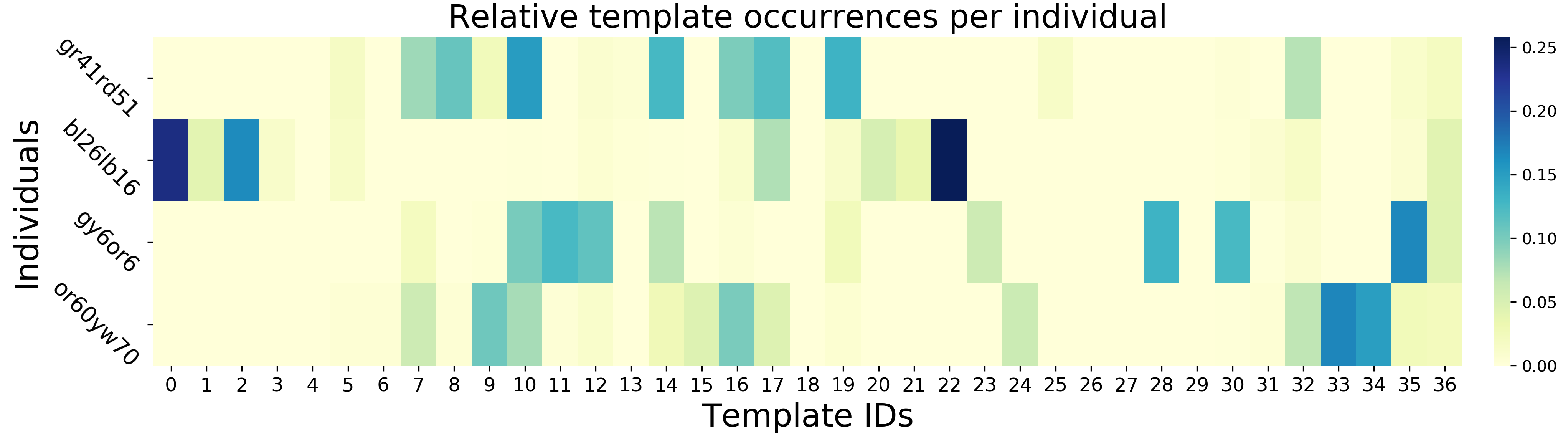}}
\caption{Relative template occurrences in each individual on the query set, using templates obtained in the multi setup.}
\label{fig:relativetemplateocc}
% \end{center}
\end{figure}

% \subsubsection*{Distinguishing individuals from their bags of syllables} 
We visualize the relative occurrences of multi-setup templates for each individual in \cref{fig:relativetemplateocc}.
Individuals share some templates but the composition of their bags of syllables differ. Moreover, certain templates are characteristic of individuals (e.g. templates 0, 2 and 22 
% are the most occurring notes in individual bl26lb16 and 
only appear in individual bl26lb16's songs).
% , and that even when individuals share some templates, the composition of their bags of syllables differ, 
This suggests that \textbf{our method can help in the task of individual identification of Bengalese finches}.

\begin{figure}[hbt!]
\centering
\centerline{\includegraphics[width=0.7\columnwidth]{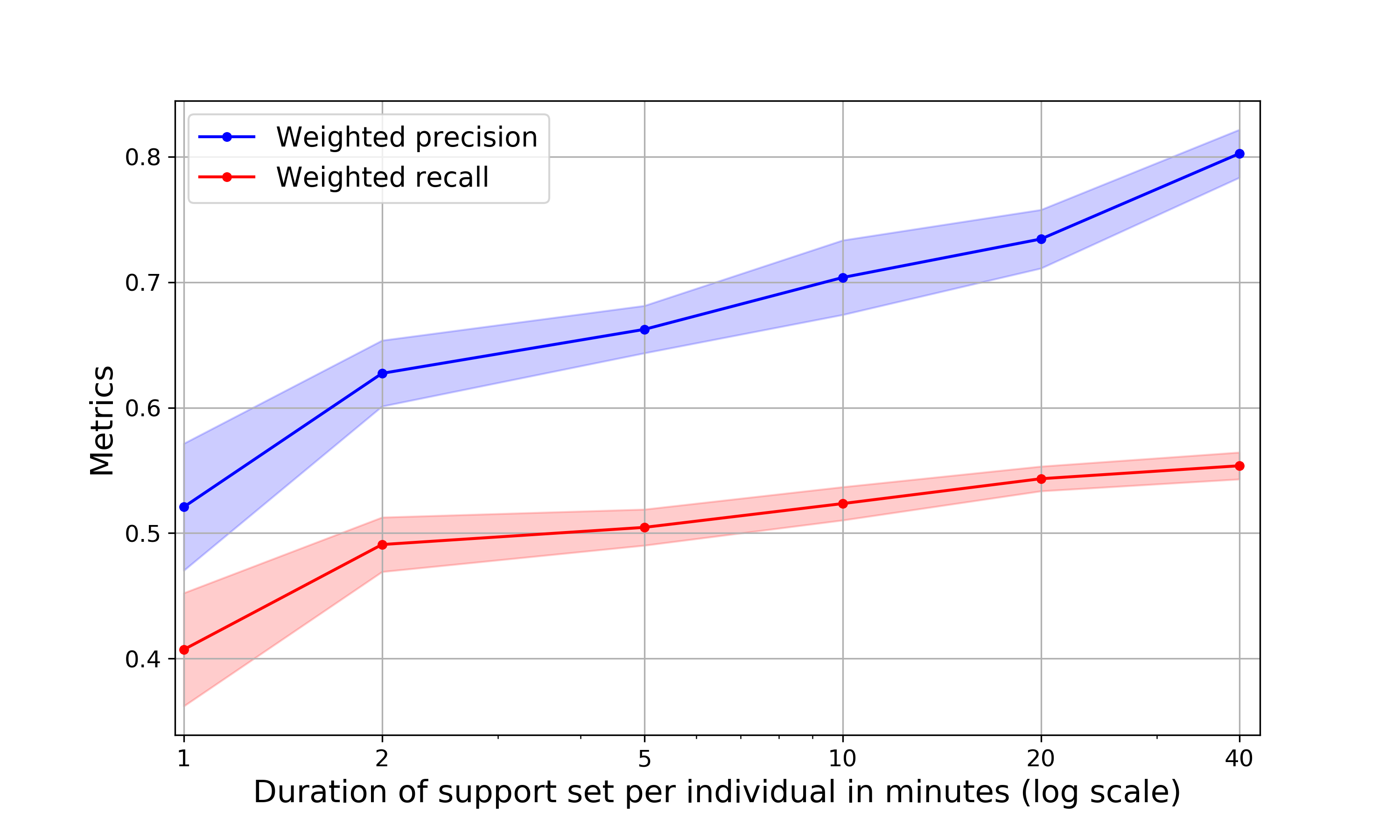}}
\caption{Weighted precision (blue) and weighted recall (red) with standard errors over 5 support/query sets, averaged over the 4 individuals, with varying duration of the support sets. } %The x-axis is log-scaled.}
\vspace{-0.25em}
\label{fig:prec-recwithtime}
% \end{center}
\end{figure}

% \subsubsection*{Additional analyses}
To assess the importance of the size of the support set, we evaluate performance as we increase the size of the support set, from 1 min to 40 min. We hold out 1 hour of recording per individual from which we sample 5 support sets using 5 random seeds per duration and individual. The query sets consist of all remaining recordings, and are thus consistent across support set sizes. We report weighted--averaged precision and recall (average and standard deviation over seeds and individuals) in \cref{fig:prec-recwithtime}. Performance increases with support set size, as templates are informed with more syllables,
% As the support set size increases, performance increases as the likelihood that all syllables in the query set are represented in a sufficient number of occurrences increases. However, 
with diminishing returns as the support set duration increases, making our experimental design choice of 10 minutes per individual a reasonable efficiency/performance tradeoff. We also found that the split-merge step helped limit oversplitting, reducing the number of clusters (and thus, templates) by a third in the multi-setup, compared to the initial clustering. 
% templates derived from the clustering at different stages of the pipeline. In our experiments, we used a single iteration of split-merge in the method and found that it improved significantly on the initial clustering

% We also report the \textbf{performance of the algorithm using templates derived from clusters at the different steps of our pipeline}. At each step, we derive templates from the current clustering, and report metrics obtained when running matching pursuit on the query sets. We use the same splits as those used in the experiments of Table \ref{tab:results}. Figure \ref{fig:steps} showcases the importance of the split-merge steps of our method, which improve significantly the templates compared to the initial clustering. 

\vspace{-0.12in}
\vspace{-0.12in}

% \begin{figure}[ht]
% \centering
% \centerline{\includegraphics[width=0.8\columnwidth]{figures/steps.png}}
% \caption{Query set weighted precision and weighted recall reported with standard deviations, on 5 random splits, averaged across the 4 individuals.}
% \label{fig:steps}
% % \end{center}
% \end{figure}
\vspace{-0.5em}

\subsection{Great Tits dataset}

\begin{figure}[hbt!]
\centering
\centerline{\includegraphics[width=0.95\columnwidth]{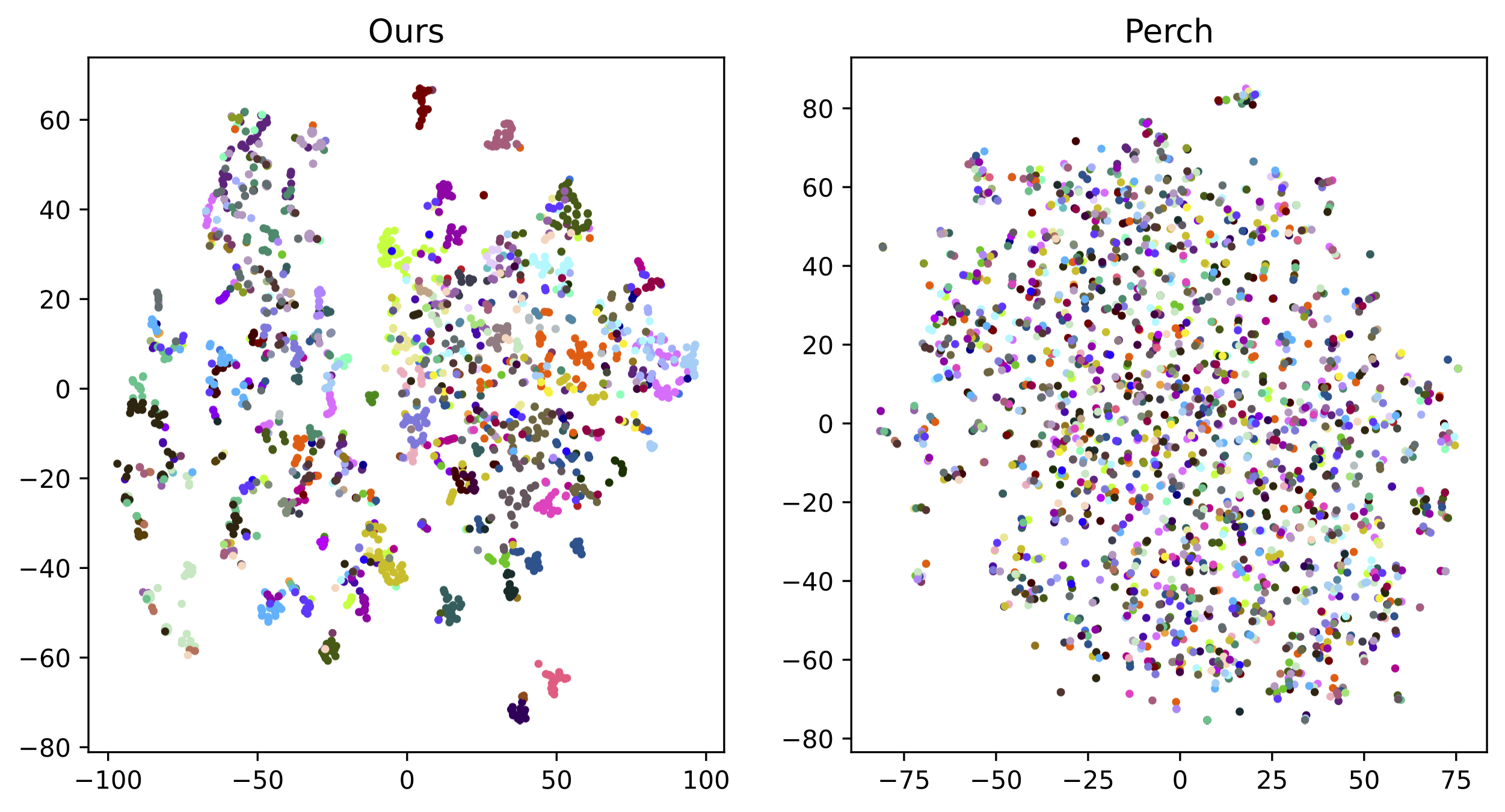}}
\caption{2D t-SNE representation of our BoS (left) and Perch embeddings (right), colored by song types and individuals.}
\label{fig:greattits}
% \end{center}
\end{figure}

Fig.~\ref{fig:greattits} shows the 2D tSNE representation of the BoS and Perch embeddings of individual songs, colored by song type and individual. We observe that our BoS embeddings encode information that separates both songs and individuals, even though no information about the order of the syllable occurrences in a sequence is used, while Perch embeddings do not. When running k-means on this 2D representation, we obtain a mean-average precision mAP $=0.46$ and a mAP@5 $=0.86$ for our embeddings vs.~mAP $=0.11$ and mAP@5 $=0.39$ for Perch embeddings, quantitatively confirming the visual impression. In total, our method finds 58 templates across all 25 individuals. The average number of individuals each template appears in at least 5\% of the time is 5.93, showing that the method finds syllables shared across individuals.  
%ALthough not using any time information. 

\vspace{-0.07in}

\section{Conclusion}
We presented a fully unsupervised method to annotate birdsongs at the syllable level. %, that can be applied to any species. 
We demonstrated promising applicability to tasks such as individual bird identification and song type clustering. We argue that our method is suited for recordings in soundproof boxes (Bengalese Finches dataset), focal recordings and clean passive acoustic monitoring data (Great Tits dataset). It may not be robust to structured noise, and the performance of our method in such cases will be investigated in future work. Further potential future directions include investigating the value of minimal human validation of the templates, and extending the method to other species such as marine mammals. %  \textcolor{blue}
%{Limitations + future directions:
%\begin{itemize}
%    \item detection when there is noise (non-white) in the recordings? if there is low SNR? - splitting by silences 
%    \item gneralizability to other species
%    \item Sequence modelling with the labelled syllables. 
%    \item human minimal refinement of templates? 
%\end{itemize}}

\vfill\pagebreak

\section{Acknowledgments}
We thank Vincent Dumoulin, Tom Denton, Yoshua Bengio, Nilo Merino Recalde, Sam Lapp, Tessa Rhinehart and members of the Kitzes lab for insightful discussions. This research was supported in part by the Canada CIFAR AI Chairs program and the Global Center on AI and Biodiversity Change (NSF OISE-2330423 and NSERC 585136). We thank the Mila IT team for their incredible support to our research community with the Mila compute infrastructure.  
%\section{COPYRIGHT FORMS}
%\label{sec:copyright}

%You must submit your fully completed, signed IEEE electronic copyright release
%form when you submit your paper. We {\bf must} have this form before your paper
%can be published in the proceedings. \cite{C2}

\bibliographystyle{IEEEbib}
\bibliography{references}

\end{document}